\documentclass{elsart}
\usepackage{graphicx, amssymb, amsfonts}
\journal{Physics Letters B}

\begin{document}

\begin{frontmatter}

\vspace*{2.0truecm}
\title{The physics of eight flavours}


\author[Groningen]{Albert Deuzeman}
\ead{a.deuzeman@rug.nl}
\author[Frascati]{Maria Paola Lombardo}
\ead{mariapaola.lombardo@lnf.infn.it}
\author[Groningen]{Elisabetta Pallante}
\ead{e.pallante@rug.nl}

\address[Groningen]{Centre for Theoretical Physics, University of Groningen, 9747 AG, Netherlands}
\address[Frascati]{INFN-Laboratori Nazionali di Frascati, I-00044, Frascati (RM), Italy}

\begin{abstract}
We study Quantum Chromodynamics with eight flavours by use of lattice simulations
and  present evidence that the theory still breaks chiral symmetry 
in the zero temperature, continuum limit. This confirms that the lower end of the conformal window of QCD lies above $N_f$ = 8.
\end{abstract}


\end{frontmatter}

\section{Introduction}
The pattern of chiral symmetry breaking in $SU(3)$ gauge theories, and more specifically QCD-like theories, with a large flavour content is not well understood. Perturbative arguments suggest that chiral symmetry should be restored above a (yet to be determined) critical flavour value $N_f^c$, and below the value $N_{AF}^c$ where asymptotic freedom disappears \cite{Banks:1981nn}. For flavour numbers between these values, the theory becomes conformal due to the presence of the Banks-Zaks infrared fixed point.

These findings were substantiated by a renormalization group estimate of the critical point \cite{Appelquist:1998rb}, and an upper bound on $N_f^c$ has been proposed \cite{Appelquist:1999hr}. For a non-supersymmetric SU(N) theory with $N_f$ massless flavours, one finds $N_f^c \le 4N\sqrt{1-16/81N^2}$. Instanton studies at large $N_f$ \cite{Velkovsky:1997fe} claimed a qualitative change in behaviour at $N_f = 6$. Renormalization group flow equations  predict $N_f^c =  10 (1)$ \cite{Gies:2005as}. Recently, an all orders beta function inspired by the Novikov-Shifman-Vainshtein-Zakharov beta function of $\mathcal{N}=1$ supersymmetric gauge theories has been computed, leading to a bound of the conformal window \cite{Ryttov:2007cx}, in particular $N_f^c>8.25$ for an SU(3) gauge group. These results have also been discussed from the more general perspective of higher dimensional representations \cite{Sannino:2005sk}.

There is more than one reason for interest in these studies. First, the uncovering of the rich phase structure of vector-like theories such as QCD, is in itself a good step forward in our understanding of gauge field theories in general. Secondly, the extent and the nature of the conformal window is of interest to model builders, since it can be used to describe electroweak symmetry breaking. Finally, the exploration of the temperature dependence of the same phenomena adds one further perspective, as there should be a connection of some sort between the conformal phase of QCD and the quark gluon plasma phase.

\begin{figure}
\center
\label{phase_sketch}
\includegraphics[width=12 truecm]{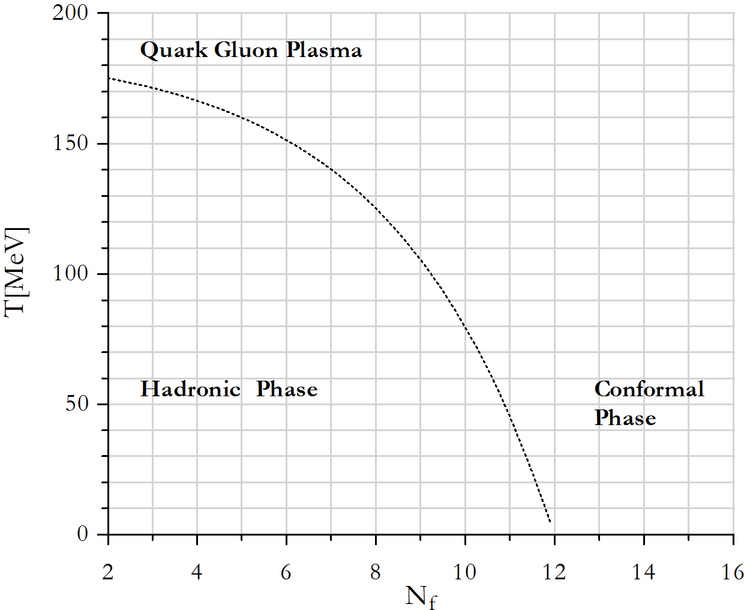}
\caption{\em{Qualitative graph of the phase diagram in the $T, N_f$ plane.} }
\end{figure}

More recently, analytical studies of QCD at large $N_f$ have been extended to finite temperature, and the critical line in the T, $N_f$ plane has been predicted on the base of a truncated renormalization group flow calculation \cite{Braun:2006jd}. Note that the number of flavours can indeed be regarded as a continuous variable, since the Casimir's are polynomials in $N_f$ \cite{Damgaard:1997ut}. This means that the critical line found in \cite{Braun:2006jd} is in fact a true phase boundary between a conventional Goldstone phase and a chirally symmetric phase, the zero temperature limit of which is the onset of the conformal phase. We sketch the resulting phase diagram in Fig.~\ref{phase_sketch}.

An interesting question is how the conformal phase of zero temperature, large $N_f$ QCD is connected with the high temperature, small $N_f$ quark gluon plasma phase. One important difference between the two is that the conformal phase possesses no inherent scale, while the other high T phase contains, of course, the temperature as a scale. As a consequence, the pion will be massless in the chiral limit of the conformal phase, while its mass will equal the lowest Matsubara frequency in the chiral limit at high T. Thus, the behaviour in the conformal window is in principle different from that in the high temperature, small $N_f$ regime of QCD. However, it is unclear if the pion mass -- or any other observable -- can be regarded as an order parameter, or, in other words, if we should envisage a further critical line at the right side of the chiral line in Fig.~\ref{phase_sketch}. Additionally, the issue whether chiral symmetry restoration, for fermions in the fundamental representation, is always accompanied by, and coinciding with, a deconfinement transition, is an open and relevant problem. Much of the numerical evidence up to date points to their entanglement, but recent work advocates a separation of the two transitions, at least in the large $N_c$ limit \cite{McLerran:2007qj,Glozman:2007tv}. It is very interesting to see what happens for large $N_f$.

Lattice calculations  provide a nonperturbative framework to attack this problem. The phase diagram of Fig.~\ref{phase_sketch} suggests a natural strategy for finding the lower limit of the conformal phase: simply, find for which $N_f$ the transition disappears. As a by-product, the numerically determined critical temperature as a function of $N_f$ can be compared with the analytic estimate of \cite{Braun:2006jd}, and the nature of
the phase transition investigated.
%
There is however one complication. Namely, it has been proposed -- and to some extent proven -- that the strong coupling limit of lattice QCD is always confining, regardless of the value of $N_f$ (see again \textit{e.g.} \cite{Damgaard:1997ut}). A simple argument is that, at large enough coupling, $N_f$ means nothing. For instance, the statement that at $N_f$ = 16 asymptotic freedom is lost, is a statement of the continuum weak coupling regime, which has no special relevance at strong coupling. Even for those $N_f$ which are supposed to fall deeply in the conformal phase, a strong coupling limit still exists which breaks chiral symmetry and confines.

This means that, even when the weak coupling continuum regime is conformal, a zero temperature `lattice' phase transition will always take place between a strong and a weak coupling regime. In other words, for a fixed number of flavours, and for a given temporal extension of the lattice $N_t$, we expect a phase transition to occur as a function of the lattice coupling to a `strongly coupled' phase where chiral symmetry is broken, i.e. the chiral condensate is non zero in the chiral limit. This type of transition, called a bulk phase transition, has to be distinguished from a true thermal transition.

We can thus envisage two types of lattice behaviour as a function of the number of flavours. When $N_f$ is such that QCD in the continuum still breaks chiral symmetry and confines, \textit{i.e.} when the continuum limit describes the ordinary phase of QCD, no transition is expected at zero temperature -- or, equivalently, $N_t= \infty$, given that $T=1/(N_t a)$, with the lattice spacing $a$ a function of the inverse lattice coupling $\beta =6/g^2$. If a lattice transition occurs it should get closer to $\beta =\infty$ and eventually disappear when $N_t\to\infty$. Instead, for the values of $N_f$ for which an infrared fixed point exists, the lattice phase transition should survive the $N_t \to \infty$ limit, \textit{i.e.} it is a bulk phase transition. The larger $N_f$, the closer this bulk phase transition should get to the infinite coupling $\beta = 0$ limit.

At a practical level, one should be able to distinguish between a bulk phase transition and a true thermal phase transition by studying the behaviour of the theory for a given $N_f$ at different values of $N_t$ and at varying lattice coupling $\beta$. If the critical point remains non zero $\beta_c \neq 0$ when $N_t\to\infty$ we have a bulk transition, thus providing a strong evidence that at the given $N_f$ the continuum theory is chirally symmetric and falls into the conformal window. The alternative behaviour will point to a thermal transition happening at a given physical temperature $T_c$.

The authors of \cite{Fukugita:1988} performed a pioneering study at $N_f=10$ and $N_f=12$ at small volumes (mainly $8^3\times 4$) and relatively heavy masses, reporting interesting evidence for a persistence of a first order thermal phase transition at $N_f=10$ and discussing its sensitivity to fermion masses. Subsequent work by the Columbia collaboration \cite{Brown:1992fz} suggested a rich structure of the phase diagram, dominated by lattice artifacts and was not completely conclusive. Iwasaki and collaborators performed a detailed lattice study, finding that a value as low as $N_f=6$ should be set as the lower limit of the conformal window \cite{Iwasaki:2003de}. Note that these authors challenge the commonly held opinion that the strong coupling limit of QCD breaks chiral symmetry, irrespectively of $N_f$. In the already mentioned paper \cite{Damgaard:1997ut}, the authors consider $N_f=16$, the largest value of $N_f$ still compatible with asymptotic freedom, and confirm that chiral symmetry is not broken at weak coupling by observing the expected bulk transition at intermediate coupling. Several other studies have observed phase transitions as a function of the gauge coupling for $N_f=8$ \cite{Kogut:1982fn, Gavai:1985wi, Kim:1992pk, Alles:2006ea, deForcrand:2007uz}, leaving the question of the fate of these transitions in the continuum limit open. Recent work by Appelquist and collaborators \cite{Appelquist:2007hu} finds evidence that $N_f = 8$ is still confining in the continuum limit. In particular, they conclude that $N^c_f$, the threshold of the conformal window, should be close to 12. They study the approach to the continuum by use of the Schr\"odinger functional and the step scaling method, the lattice discretized version of the renormalization group running coupling, and do not discuss how a bulk transition would affect these calculations.

In summary, most of older studies at $N_f=8$ suggested that the theory is already in the conformal window, while others confirmed analytic calculations \cite{Banks:1981nn, Appelquist:1998rb, Ryttov:2007cx, Braun:2006jd}, indicating that $N_f^c > 8$, close to the upper limit for $N_f^c$ calculated in \cite{Appelquist:1999hr}.

The purpose of this study is to fully clarify the physics of eight flavours, as a first step towards the exploration of the phase diagram of QCD at large $N_f$.

\section{Simulation and Observables}

The simulations in this paper were performed using a slightly modified version of the publicly available MILC code \cite{MILCwebsite}. A setup similar to the one used by MILC in their recent paper on the QCD equation of state \cite{Bernard:2006nj} was employed.

We use an improved Kogut-Susskind fermion action, the ``Asqtad'' action which removes lattice artifacts up to $O(a^2 g^2)$ \cite{MILCwebsite} and a one-loop Symanzik improved \cite{Bernard:2006nj, LW_1985} and tadpole improved \cite{LM_1985} gauge action. The complete action for $N_f$ mass degenerate flavours can be written as
\begin{equation}
\label{eq:action}
S=  - \frac{N_f}{4}\, {\mathrm{Tr}}\ln{M(am,U, u_0)} + S_{gauge}\, ,
\end{equation}
where
\begin{equation}
S_{gauge} = \sum_{i=p,r,pg}\beta_i(g^2) \sum_{  {\cal{C}}\in  {\cal{S}}_i } Re (1- U ( {\cal{C}} ))\, ,
\nonumber
\end{equation}
with couplings defined as
\begin{eqnarray}
\beta_p &\equiv& \beta = 10/g_0^2 \nonumber\\
\beta_r &=& -\frac{\beta}{20u_0^2}(1+0.4805\alpha_s)\nonumber\\
\beta_{pg} &=& -\frac{\beta}{u_0^2}0.03325\alpha_s\, ,\nonumber\\
\end{eqnarray}
with $M(am, U, u_0)$ the fermion matrix for the Asqtad staggered action for a single flavour with mass $m$, and with $\alpha_s = -4\log{u_0}/3.0684$. 
The ${\cal{S}}_i$'s contain all the $1\times 1$ plaquettes, the $1\times 2$ and $2\times 1$ rectangles and the $1\times 1\times 1$ parallelograms, respectively, that can be drawn on the lattice. The $U ({\cal{C}})$'s are the traces of the ordered product of link variables along ${\cal{C}}$, all divided by the number of colours. The tadpole parameter $u_0$ is defined in terms of the gauge invariant average plaquette as $u_0 = \langle U ({\cal{C}}) \rangle |_{ {\cal{C}}\in {\cal{S}}_p }^{1/4}$. Exploiting the rather low  sensitivity of the plaquette value to finite volume effects, the $u_0$ values for each run were determined by a self consistency procedure on $12^4$ and $16^4$ lattices for every $\beta$ investigated. 

It has been proposed recently by members of the HPQCD collaboration \cite{Hao:2007iz} to include the effect of dynamical quarks into the one-loop improvement of the gauge action, the required coefficients of which they determined to first order through lattice perturbation theory. While it is true that a complete $O(a^2)$ improvement should fully account for gauge and quark loop contributions, the reliability of a one-loop truncated 
perturbative contribution from dynamical quarks at large $N_f$ becomes questionable. Since the coefficients found in \cite{Hao:2007iz} are sizable and linear in the flavour number, a resummation would probably be called for, especially in the case of large $N_f$. Barring this, there is a risk that including a truncated series would severely overcompensate the actual quark loop effects and in fact worsen the $O(a^2)$ improvement, for values of $N_f \geq 3$. The scaling violations that are being observed for the heavy quark potential with the customary `partially' improved action are not completely negligible, but well under control \cite{Davies:2008pc}. On the base of these considerations and the fact that observables employed in this study are less sensitive to scaling violation effects, we did not include these one-loop dynamical quark improvements.

To generate configurations with eight degenerate dynamical flavours, we used the rational hybrid monte carlo (RHMC) algorithm \cite{Clark:2006wq}, which allows for simulating an arbitrary number of flavours through varying the number of pseudofermions used, \emph{i.e.} it inherently uses an analytical continuation in the number of flavours. Simulations were done with two pseudofermions, thus avoiding fractional powers of the Dirac operator in the molecular dynamics, and actually maintaining the best performance if compared with other choices of the number of pseudofermions. We are fully exploiting the characteristics of the RHMC algorithm and monitor possible effects of a rooting procedure -- by the use of fractional powers of the Dirac operator -- in ongoing studies at varying number of flavours \cite{DLP:Conformal}.

To systematically investigate the character of eight flavour QCD under different thermodynamical conditions, simulations were run at two different values for the length of the temporal extent in lattice units $N_t$, namely 6 and 12. As was mentioned in the introduction, the scaling behaviour of the critical lattice coupling, defined according to eq.~(\ref{eq:action}) as $\beta_c = 10/g_{0c}^2$, with this parameter is an indication of either the thermodynamical or bulk nature of the observed phase transition. To check for the influence of finite volume effects and deduce the scaling properties of observables, simulations were run for three spatial extents of the lattice $N_s = 12, 20, 24$ for $N_t = 6$, and $N_s=24$ for $N_t=12$. 
We chose a fixed value of the lattice degenerate quark mass $am = 0.02$ to explore the critical region, and will describe how results can be modified by doing simulations at a lower quark mass in section \ref{sec:Nt12}. As it will be clear from the outcomes, the chosen masses are sufficiently light to clearly disentangle the critical behaviour and establish the nature of the transition.

The Monte Carlo history was collected with trajectories of total length $\tau =0.3$ to  $0.4$, and time step $\delta \tau = 0.003$ to $0.007$ from the lowest to the highest $\beta$ values and from the smallest to the largest volume.

At each $\beta$, the expectation value of the real part of the Polyakov loop 
\begin{equation}
L \equiv \frac{1}{3 N_s^3}\,\sum_{\vec{x}}\mathrm{Re}\, \mathrm{Tr} \prod_{x_4=1}^{N_t} U_4({\vec{x}}, x_4)
\end{equation}
was determined. It is important to remember that the real part of the Polyakov loop is a true order parameter only for the pure gauge theory, which is recovered in the infinitely heavy mass limit $m\to\infty$ of the theory with dynamical flavours. It is not an order parameter in all other cases, though a clear change in its value and its susceptibility can be observed at sufficiently heavy or sufficiently light values of the fermion masses. Where the theory does enter the regime of `light masses' is based on empirical observations \cite{Nf4_PRL}, it might shed light on the mechanism relating confinement to chiral symmetry breaking, and it depends on the flavour content of the theory.

The chiral condensate for $N_f$ degenerate flavours in lattice units
\begin{equation}
a^3\left\langle \bar{\psi} \psi \right\rangle = \frac{N_f}{4N_s^3N_t}\langle \mathrm{Tr} \left[M^{-1}\right] \rangle \, ,
\end{equation}
was determined by using a stochastic estimator with 20 repetitions. The chiral susceptibility, measuring the variation of the chiral condensate with varying the fermion mass $\chi = \partial \langle \bar{\psi} \psi \rangle /\partial m$ at fixed $\beta$ can be divided into a connected and disconnected component $\chi = \chi_{\mathrm{conn}}+\chi_{\mathrm{disc}}$, given in lattice units by 
\begin{eqnarray}
a^2\chi_{\mathrm{conn}} &=&  -\frac{N_f}{4 N_s^3 N_t}   \langle \mathrm{Tr} \left[( MM )^{-1}\right ] \rangle                  \nonumber\\
a^2\chi_{\mathrm{disc}} &=& \frac{N_f^2}{16 N_s^3 N_t}\left [  \langle \mathrm{Tr} \left[M^{-1}\right] ^2\rangle - \langle \mathrm{Tr} \left[M^{-1}\right] \rangle^2  \right ]\, ,
\end{eqnarray}
respectively. We have conveniently written the condensate and its susceptibilities in terms of traces of (products of) the staggered fermion matrix $M$ as they are actually computed in the simulation. The connected and disconnected contributions to the chiral susceptibility are measured separately; more on this can be found in \cite{Bernard:1996zw}. The connected contribution can also be measured in a partially quenched manner, performing a numerical derivative of the chiral condensate with respect to the valence quark mass
\begin{equation}
\chi_{\mathrm{conn}} = \frac{\partial \langle \bar{\psi}\psi \rangle}{\partial m_V}\, .
\end{equation}
Both measurement methods were implemented, with results being in excellent agreement. The complete set of values reported on here have been found using the first method, since it has a lower computational overhead.

The disconnected chiral susceptibility is a non-local quantity that can be estimated from the variance of the bulk behaviour of the chiral condensate. However, such a variance for a condensate computed with stochastic estimators will automatically include part of the connected contributions, through random sources multiplying themselves. Following Bernard \textit{et al.} \cite{Bernard:1996zw}, we straightforwardly eliminate those contributions by only considering the off-diagonal elements of the covariance matrix of the random sources introduced for the estimation of the chiral condensate.

We can use the chiral susceptibility and the chiral condensate to define two physically relevant quantities
\begin{equation}
\chi_\sigma \equiv\chi = \frac {\partial \langle \bar \psi \psi\rangle}{\partial m} = \chi_\mathrm{conn} + \chi_\mathrm{disc}
\end{equation}
and
\begin{equation}
\chi_\pi = \frac {\langle \bar \psi \psi\rangle}{m}\, .
\end{equation}
They are related through Ward identities \cite{Kocic:1992is} to the  spacetime volume integral of the 
scalar ($\sigma$) and pseudoscalar ($\pi$) propagators
\begin{equation}
\chi_{\sigma ,\pi} =  \int ~d^4x~G_{\sigma,\pi } (x)\, ,
\end{equation}
thus implying that they should become degenerate when chiral symmetry is restored, following the degeneracy of the chiral partners. As a result, their associated cumulant $R_\pi \equiv \chi_\sigma/\chi_\pi$ should be, in the absence of explicit chiral symmetry breaking, one in a chirally symmetric regime, while it should be zero in the spontaneously broken phase. This dimensionless quantity is therefore a most useful physical observable in analyzing chiral phase transitions, and we will employ it to compare results from simulations run at different temporal extents of the lattice.

\section{Results at $N_t=6$}

The introductory discussion tells us that we must observe a phase transition between a phase where chiral symmetry is broken to a phase where chiral symmetry is realized, for some value of the lattice gauge coupling $\beta_c$. There are basically two ways to assess a phase transition on a lattice. First, one can rely on obtaining infinite volume estimates of relevant quantities by performing simulations on lattices of different spatial extents. Whenever this is possible, the critical behaviour is analogous to the one observed in the continuum. This strategy usually breaks down very close to the critical point, where the correlation length of the system grows large and becomes comparable to the lattice size. When this happens, it is mandatory to use finite size scaling techniques. The determination of the regime we are in can be made \textit{a posteriori}, by comparing results obtained from different volumes.

\begin{figure}
\center
\includegraphics[width=12 truecm]{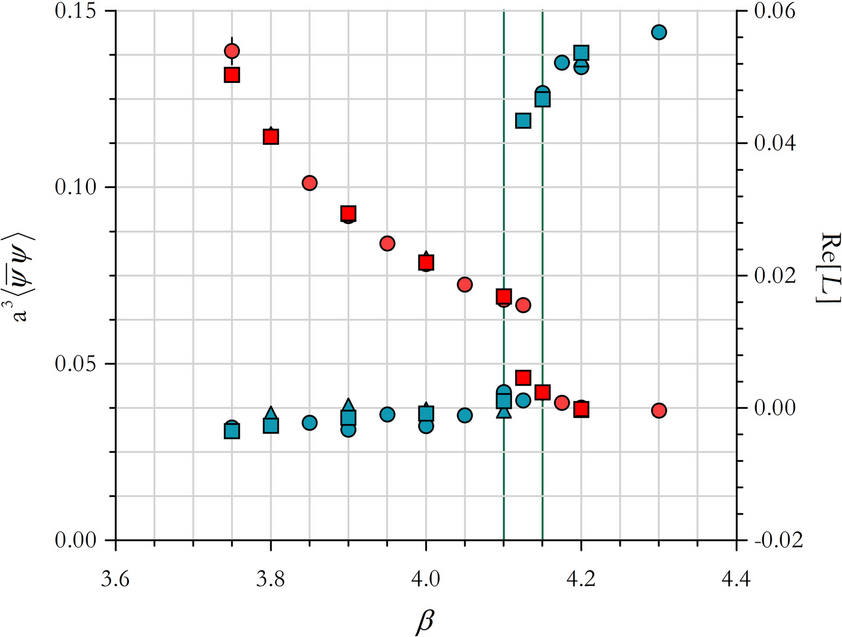}
\caption{\it The chiral condensate (red) and Polyakov loop (blue) as a function of the lattice coupling $\beta$. Measurements were done for three different spatial extents of the lattice $N_s$: 12 ($\bigcirc$), 20 ($\triangle$) and 24 ($\Box$). The critical region has been indicated by vertical lines.}
\label{fig:pbp_pol_nt6}
\end{figure}

For the case at hand, results for the chiral condensate and the Polyakov loop are shown in Fig. \ref{fig:pbp_pol_nt6}. From the small differences found at different spatial extents, it was concluded that our results can indeed be considered infinite volume estimates for $\beta \le 4.1$ and $\beta \ge 4.15$, and the $\beta$ dependence is smooth. The jump between the two branches is very clear for both observables and suggestive of a discontinuity.

These results allow us to immediately identify a critical region for $\beta$ between 4.1 and 4.15. In this region, one would expect to observe tunnelling between the two phases and enhancement of the associated susceptibilities. As mentioned above, the approach to the infinite volume limit should be slower here, because of the increasing correlation lengths. Indeed, volume dependence is observed, with the smaller of the 
lattices having a slightly higher $\beta_c$. This behaviour is consistent with known predictions for finite volume effects on first order phase transitions \cite{Karsch:1989pn}. Since the shift is much less pronounced for the change of volume between $N_s=20$ and $24$, we concluded the infinite volume limit value for $\beta_c$ should be located at the lower parts of the designated critical region, and is bounded by $4.10 < \beta_c < 4.125$. For definiteness, we locate it at a value of 4.1125.

\begin{figure}
\center
\includegraphics[width=12 truecm]{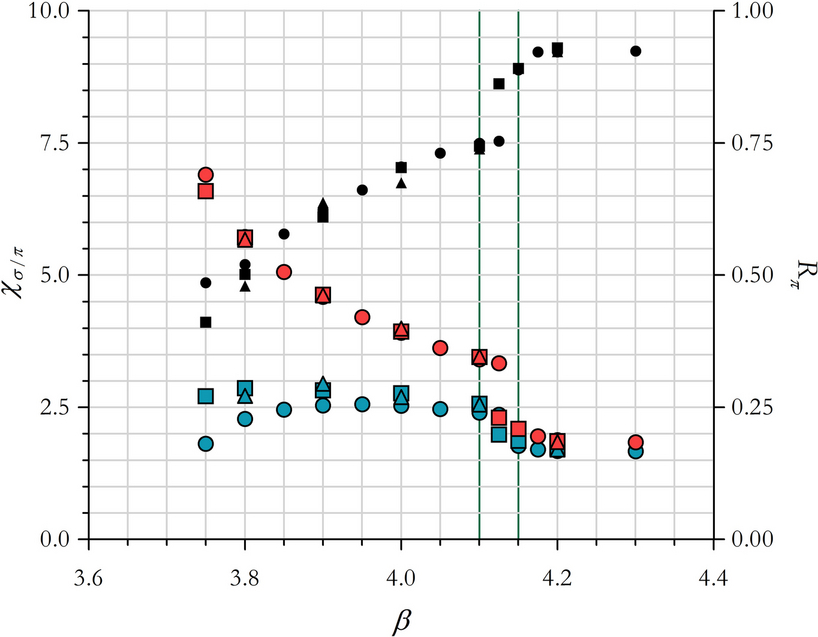}
\caption{\it The scalar $\chi_\sigma$ (blue) and pseudoscalar $\chi_\pi$ (red) chiral susceptibilities
as a function of the lattice coupling $\beta$, confirming the degeneracy of
the chiral partners in the symmetric phase. Measurements are shown for three different spatial extents of the lattice $N_s$: 12 ($\bigcirc$), 20 ($\triangle$) and 24($\Box$).The cumulant $R_\pi$ is also shown (black, top of figure). The critical region has been indicated by vertical lines.}
\label{fig:chisusc_nt6}
\end{figure}

The chiral susceptibilities are shown in Fig. \ref{fig:chisusc_nt6} and confirm our picture. The residual splitting is associated with the explicit breaking induced by the mass term, but the tendency towards degeneracy is very clear. Finally, we plot the cumulant $R_\pi$, defined earlier, in Fig. \ref{fig:chisusc_nt6}. In the chiral limit, the cumulant should be zero in the broken phase, and approach one in the chirally symmetric phase. The observed trend is consistent with all other findings.

\subsection{The critical region}

\begin{figure}
\center
\includegraphics[width=12 truecm]{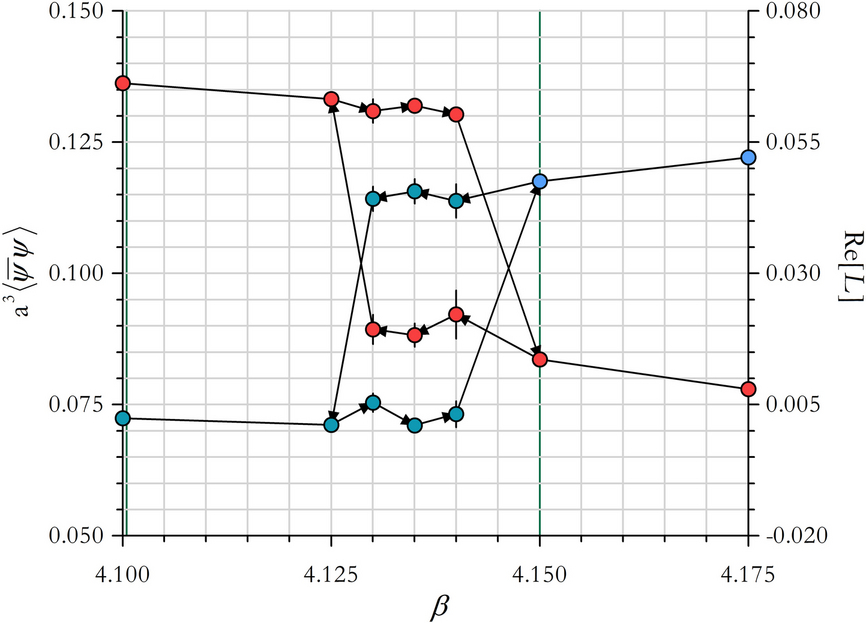}
\caption{\it Hysteresis effect for the chiral condensate (red) and Polyakov loop (blue) within the critical region, data are for the spatial extent $N_s=12$.
Arrows indicate the $\beta$ of the configurations involved, the vertical lines indicate the estimate of the broadest critical region which covers the jump of the chiral condensate and Polyakov loop at all spatial volumes and the hysteresis cycle.}
\label{fig:hysteresis_loop}
\end{figure}

Having bracketed off the critical value for $\beta$, we investigated the area in which the phase transition occurs in more detail. Of specific interest here is the potential presence of metastabilities. Simulations were performed on $12^3\times6$ lattices starting from thermalized configurations at higher and lower values of $\beta$, but with identical parameters. The presence of metastable states could show up, either through the presence of hysteresis effects, or by appearance of tunnelling effects in the Monte Carlo history. The latter effect would be expected to be somewhat enhanced, because of the usage of a smaller spatial extent. However, for all of the values of $\beta$ that were examined near the threshold of the lattice of this size, a persistence of initial states was found after a large number of RHMC trajectories. This indicates the presence of a strong hysteresis loop with an extent equal to about half of our designated critical region, the consequences of which for the main observables have been plotted in Fig. \ref{fig:hysteresis_loop}.

\begin{figure}
\center
\includegraphics[width=10 truecm]{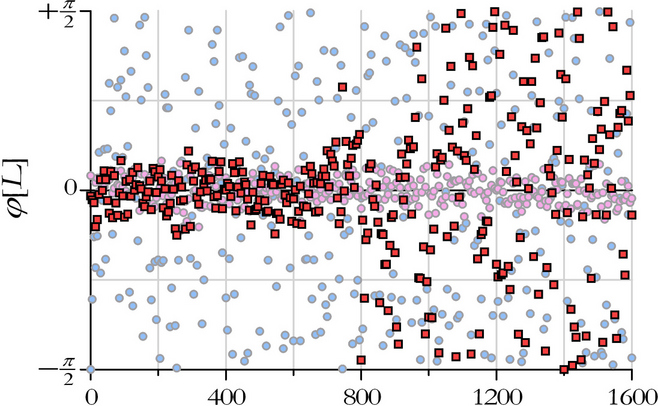}
\caption{\it Monte Carlo history for $\beta=4.1$ and $N_s=24$, at the threshold. The distribution flattens around zero in the deconfined phase, while it is randomly distributed between $\pm\frac{\pi}{2}$ in the confined phase. The same histories are shown for runs below and above the critical region (at $\beta=4.0$ and $4.2$) in lighter colours. The points are enumerated according to their trajectory number, disregarding the first 100 trajectories after the cold start for each.}
\label{fig:ploop_phase}
\end{figure}

Finally, it is a matter of interest to what extent the behaviour of the Polyakov loop, as a signal of confinement, is intertwined with that of the chiral condensate. There are indications that the confinement/deconfinement transition is, in the presence of light quarks, driven by the chiral phase transition. Even though the signal of the Polyakov loop tends to be statistically weak for small numbers of measurements, it is possible to observe its change from the distribution of its phase. While the Polyakov loop tends to be distributed symmetrically in the complex plane for confining configurations, whenever fermions are present in a deconfined system, it will tend towards the real phase of its $Z(3)$ centre. In Fig. \ref{fig:ploop_phase}, we exhibit the Monte Carlo history of the phase of the Polyakov loop for a run at the threshold of the critical region and at the largest spatial extent, \textit{i.e.} $\beta=4.1$ and $N_s=24$. For comparison, the plot includes the equivalent part of the history at values of $\beta$ before and after the transition. Note the presence of a metastability close to the threshold, also for this largest volume.  This shows that the Polyakov loop, though it is a true order parameter only in the pure gauge theory, retains critical behaviour close to the transition, even in the presence of eight degenerate flavours.

All data at $N_t=6$ point towards a first order phase transition. The location of the jump of the chiral condensate and the Polyakov loop at all three spatial volumes considered, and the location of the hysteresis cycle at the smallest spatial extent, allows us to give the estimate of the broadest critical region, to be $ 4.10 <\beta_c <4.15$. We take the occurrence of the jump of the chiral condensate and the Polyakov loop  at our largest spatial volume $N_s=24$, as the best indicator of the critical point, giving $\beta_c (N_t = 6, N_s = \infty )  = 4.1125 \pm 0.0125$, corresponding to the upper and lower bound $4.10 <\beta_c (N_t=6, N_s = \infty ) <4.125$.

\section{Results at $N_t=12$}
\label{sec:Nt12}

As explained in the introduction, the scaling of $\beta_c$ with varying the lattice temporal extension $N_t$ should distinguish between a thermal and a bulk transition. A thermal transition occurs at a physical temperature $T_c$ which we determine as 
\begin{equation}
\label{eq:Tc}
T_c = \frac{1}{a(\beta_c ) N_t}
\end{equation}
for a given temporal extent $N_t$ and critical lattice spacing $a (\beta_c)$. Thus, performing simulations at different $N_t$, the critical coupling should rescale in order to give the same critical temperature $T_c$. Uncertainties in the determination of the critical temperature can be mainly introduced by lattice violations to the asymptotic scaling of observables. It is thus a possibility that past simulations leading to apparent scaling violations and indicating the presence of a bulk transition,  might have been performed in a region of the lattice parameter space not sufficiently close to the continuum limit. Our highly improved lattice action should help us approaching the continuum limit and the correct scaling behaviour.

\begin{figure}
\center
\includegraphics[width=12 truecm]{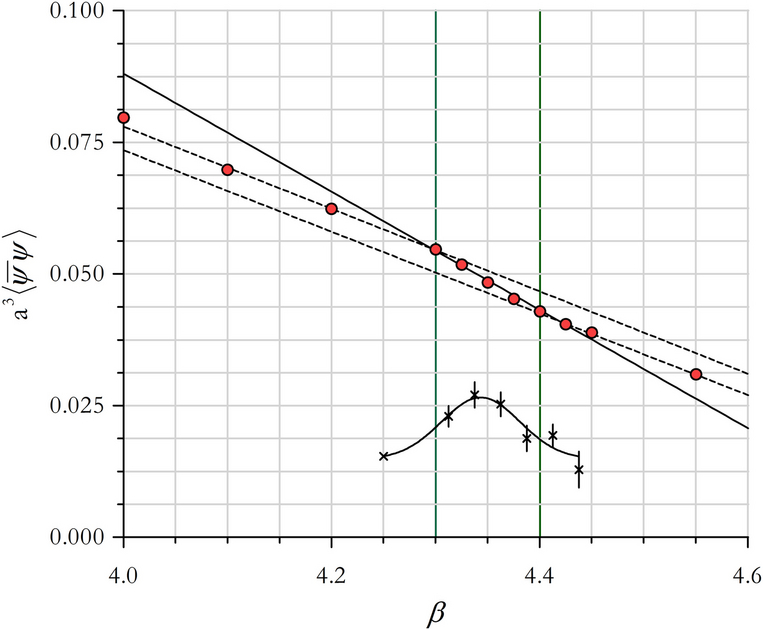}
\caption{\em{The chiral condensate at $N_t=12$ and $N_s=24$ in lattice units  as a function of the lattice coupling $\beta$. Best fit curves are superimposed and the vertical lines indicate the critical region. The absolute value of the finite difference between measured values of the condensate, as an approximation of its first derivative, is plotted in the bottom part of the figure with an arbitrary rescaling. It shows a peak at $\beta = 4.34$.}}
\label{fig:pbp_Nt12}
\end{figure}

We have thus performed simulations at $N_t = 12$ with spatial volume $N_s=24$ and at the same lattice mass $am=0.02$, and looked for signals of the transition in the behaviour of the chiral condensate. The main result is shown in Fig. \ref{fig:pbp_Nt12}, where we observe a jump in the chiral condensate suggestive of a discontinuity, and slightly distorted and smoothed by the not so large spatial extent $N_s=24$ as compared with the temporal extent $N_t=12$. Notice also that effects of explicit chiral symmetry breaking due to the non zero physical mass are to be more pronounced at $N_t=12$, for the same value of the lattice mass. The derivative of the condensate is also plotted in Fig. \ref{fig:pbp_Nt12} to better locate the transition. The best fit value for the maximum of the derivative indicates a value for the critical coupling $\beta_c = 4.34\pm 0.04$. This result allows us to place an upper and lower bound on the critical coupling $\beta$: $4.30 <\beta_c <4.38$ at $N_t=12$, with $N_s=24$. While finite spatial volume effects will not effect our conclusion, we cannot quote this value as the infinite volume value, and we postpone a more refined analysis of finite volume effects to a future publication.

The combined results at $N_t=12$ and $N_t=6$ strongly suggest the occurrence of a first order thermal transition. To confirm this, an asymptotic scaling analysis is needed in order to verify that we are actually measuring a critical temperature in the continuum real world. We do so by means of the standard relation that connects the lattice cutoff $\Lambda_L$ to the gauge coupling $g$, $a\Lambda_L = R(g^2 )$ with 
\begin{equation}
\label{eq:2loop}
R(g^2 ) =    (b_0 g^2)^{-{b_1}/{2b_0^2}}\, e^{-1/2b_0 g^2}\, ,
\end{equation}
where the two loop RG running of the $\beta$-function is accounted for, with the universal one- and two-loop coefficients given by 
\begin{eqnarray}
b_0 &=& \frac{1}{16\pi^2}\left ( 11-\frac{2}{3} N_f\right ) \nonumber\\
 b_1 &=& \frac{1}{(16\pi^2)^2}\left ( 102-\frac{38}{3} N_f\right )
\end{eqnarray}
for $N_f$ massless flavours.Given a physical temperature $T_c$, eq.~(\ref{eq:Tc}) implies the scaling relation $N_t\, R(g_c (N_t)) = \mathrm{const}$. Solving for $N_t =6$ and $N_t=12$, we can predict $\beta_c (g_c(N_t=6))$ by knowing $\beta_c (g_c(N_t=12))$. A strong discrepancy with the actual lattice determination might besuggestive of a bulk zero temperature transition, while a small discrepancy can indeed be expected and imputed to violations of asymptotic scaling and residual effects due to a non zero fermion mass. The question is what coupling $\beta_c(g_c)$ is the most appropriate one to insert in eq.~(\ref{eq:2loop}) in order to compare continuum perturbation theory with lattice calculations. Our simulations use a one-loop Symanzik improved and tadpole improved gauge action. These improvements are expected to ameliorate the agreement with the asymptotic scaling by correcting for up to and including $O(a^2)$ effects. In a finite temperature study this translates into the statement that the agreement with asymptotic scaling will set in at lower values of $N_t$ with respect to an unimproved lattice action.

Given these premises, we have predicted the critical coupling $\beta_c (N_t=6)$ by using an effective coupling which relates the lattice $\beta$ and $g^2$ as $\beta = 6/g^2$ in eq.~(\ref{eq:2loop}). Notice that the lattice bare coupling, coefficient of the plaquette action in ~eq. (\ref{eq:action}), is $\beta_{pl} = 10/g_0^2$ where the rescaling of the coefficients is due to the one-loop Symanzik improvement. The use of $\beta = 6/g^2$ in the RG formula is meant to effectively take into account the tadpole and Symanzik one-loop improvement in our action.

By using $\beta_c (N_t=12) =4.34\pm 0.04$ we obtain the prediction $\beta_c (N_t=6)= 4.04 \pm 0.04$, which deviates by less than $2\%$ from the lattice determined $\beta_c$. We also verified that a rescaling of the effective coupling $\beta \to \beta u_0^{-4}$ \cite{LM_1985} improves the prediction by only $0.5\%$ indicating that our effective coupling prescription is in fact already accounting for all the improvement. Further improvement of the lattice gauge action -- e.g. the proper inclusion of quark loop corrections in the improvement program -- and corrections for the non zero fermion mass, should account for the tiny residual asymptotic scaling violations in this study.

\begin{figure}
\center
\includegraphics[width=12 truecm]{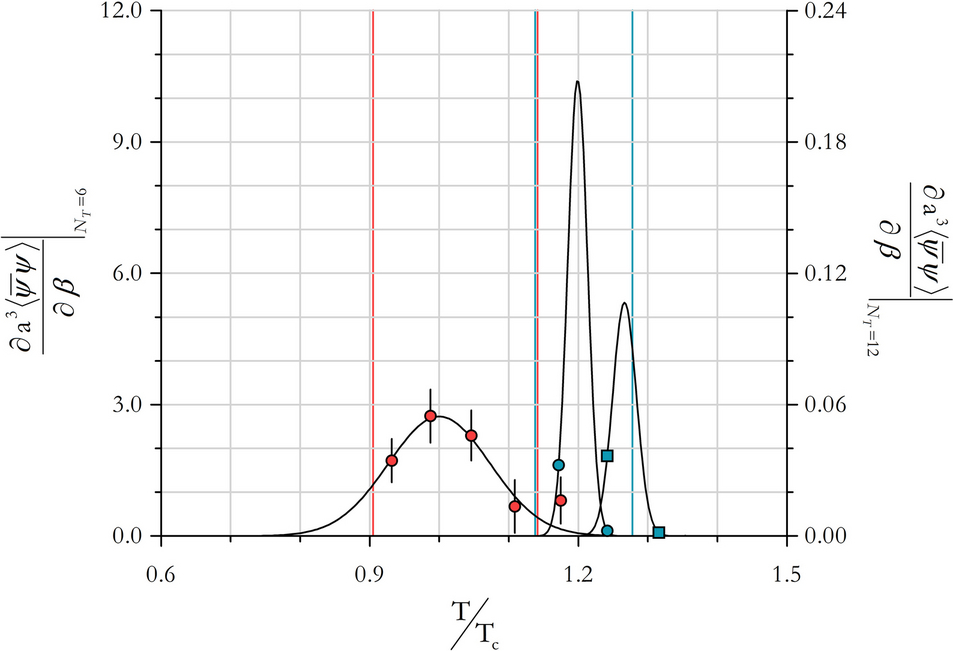}
\caption{\em{`Scaling plot' for the finite difference approximation to the absolute value of the first derivative of the chiral condensate as a function of $T/T_c$, determined using the effective coupling $\beta =6/g^2$ in the RG formula. Data at $N_t=6,\, N_s=24, 12$ (blue) are compared with data at $N_t=12,\, N_s=24$ (red). The point $T/T_c=1$ corresponds to $\beta_c(N_t=12) = 4.34$. A gaussian fit is superimposed to the $N_t=12$ data, while indicative gaussian curves are shown for $N_t=6$. The baseline is subtracted and we only indicate data near the critical points. The lattice determined critical regions at $N_t=6$ (blue) and $N_t=12$ (red) are indicated by vertical lines.}}
\label{fig:scaling_r}
\end{figure}

In Fig. \ref{fig:scaling_r} we compare the transition at $N_t=6$ and at $N_t=12$ by means of the first derivative of the chiral condensate as a function of $T/T_c$ as predicted by the asymptotic scaling using the effective coupling $\beta=6/g^2$. As before, the lattice determined $\beta_c(N_t=12)$ corresponds to the point $T/T_c=1$. The most important observation is that the transition at $N_t=6$ is close to the predicted point $T/T_c=1$, actually happening at $T/T_c \simeq 1.20$. We show the data of the first derivative with a gaussian fit superimposed. For the data 
at $N_t=6$ we show, for the sake of comparison, the curves at the largest spatial extent $N_s=24$ and the smallest $N_s=12$. The lattice determined critical regions at $N_t=6$ and $N_t=12$ overlap at their boundaries. The amount of non-overlap is a measurement of the residual but small violations of asymptotic scaling.
We did not observe any signal of discontinuity in the Polyakov loop in our simulations at $N_t=12$, with a lattice mass $am=0.02$ and in the critical region of the chiral condensate. While the presence or absence of this signal is of no physical relevance below the infinite mass limit, given that 
the Polyakov loop is not an order parameter in the presence of dynamical fermions, we can still use its behaviour at light quark masses to gain further insight into the mechanism that relates chiral symmetry restoration to deconfinement and the two limiting regimes of zero mass and infinite mass fermions. Here, we limited ourselves to verify that the presence or absence of a discontinuity of the Polyakov loop in coincidence with the discontinuity of the chiral condensate depends upon the physical fermion mass. By sufficiently lowering the fermion mass at $N_t=12$ 
we expect that the picture of $N_t=6$  will be restored.

However, the region of sufficiently light quark mass is a dynamical property of the $N_f =8$ theory which cannot be judged a priori. A further refinement of this study can help in locating the critical end-points.  The reader might want to review \cite{Nf4_PRL} for a pioneering study at $N_f=4$. We postpone to future publications a more refined study of the quark mass dependence of the chiral condensate, and a thorough scale setting procedure for the results obtained in this work.

The present results allow us to firmly conclude that we are seeing a true thermal transition, in other words $N_f=8$ undergoes a chiral restoration transition at finite temperature.

\section{Discussion}

We find $\beta_c (N_f=8, N_t=6, N_s = \infty) = 4.1125 \pm 0.0125$, or $4.1\geq\beta_c (N_f=8, N_t=6, N_s = \infty) \leq 4.125$
and $\beta_c (N_f=8, N_t=12, N_s = 24) = 4.34\pm 0.04$, or $4.30\geq\beta_c (N_f=8, N_t=12, N_s = 24) \leq 4.38$ and observe the asymptotic scaling of the critical temperature within $20\%$ by use of the perturbative scaling induced by asymptotic freedom, where the two loop beta function for an $SU(3)$ gauge theory with $N_f=8$ massless flavours has been used. These observations, obtained with an improved lattice action, might well explain the lack of asymptotic scaling noted in previous studies, where no improvement was yet considered.

All evidence presented in this paper is consistent with asymptotic scaling and the occurrence of a true thermal transition. The emerging picture confirms the conclusion of the recent work \cite{Appelquist:2007hu}, as well as ladder calculations \cite{Braun:2006jd}, providing a strong evidence that $SU(3)$ gauge theory with eight flavours is in the normal, chirally broken phase of QCD at zero temperature and in the continuum limit. This study is one step in our attempt to clarify the way $SU(3)$ gauge theory with many flavours approaches the conformal phase \cite{DLP:Conformal} that precedes the loss of asymptotic freedom.

\section*{Acknowledgements}

This work was in part based on the MILC collaboration's public lattice gauge theory code. See http://physics.utah.edu/$\sim$detar/milc.html for details. The authors  would like to acknowledge the many useful discussions and correspondence with Philippe de Forcrand, Rajiv Gavai, Holger Gies, Vladimir Miransky, Enrico Nardi, Francesco Sannino, Doug Toussaint, Koichi Yamawaki  and in particular Carleton de Tar. MPL wishes to  thank the Centre for Theoretical Physics in Groningen for its hospitality while this work was started, as well as the Institute for Nuclear Physics at the University of Washington and the U.S. Department of Energy for partial support. These simulations were performed on the Stella IBM BG/L of the Rekencentrum of the University of Groningen and Astron, as well as on the Huygens IBM Power5+ system located at Stichting Academisch Rekencentrum Amsterdam. The latter was supported by grant nr. SH-079-08 of the Dutch Nationale Computer Faciliteiten (NCF) foundation.

\end{document}